# Spin-Enhanced Organic Bulk Heterojunction Photovoltaic Solar Cells

Ye Zhang,[1] Tek P. Basel,[2] Bhoj R. Gautam,[2] Xiaomei Yang,[1]
Debra J. Mascaro,[3] Feng Liu,[1] and Z. Valy Vardeny [2*]

[1] *Department of Materials Science & Engineering,* [2] *Department of Physics & Astronomy,* [3] *Department of Mechanical Engineering*
*University of Utah, Salt Lake City, Utah 84112*

Recently much effort has been devoted to improve the efficiency of organic photovoltaic (OPV) solar cells based on blends of donors (D) and acceptors (A) molecules in bulk heterojunction architecture. One of the major losses in OPV devices has been recombination of polaron pairs (PP) at the D-A domain interfaces. Here we discovered a novel method to suppress PP recombination at the D-A domain interfaces and thus improve the OPV solar cell efficiency, by doping the device active layer with spin ½ radical Galvinoxyl. At an optimal doping level of 3 wt%, the efficiency of a standard P3HT/PCBM solar cell improves by 18%. A spin-flip mechanism is proposed and supported by magneto-photocurrent measurements, as well as by density functional theory calculations in which PP recombination rate is suppressed by resonant exchange interaction between the spin ½ radicals and charged acceptors, which converts the PP spin state from singlet to triplet.

## Introduction

Solar energy has been identified as the leading renewable energy source to meet the challenge of increasing energy demand. Organic photovoltaic (OPV)[1-25] is an emerging sector in the photovoltaic industry that has seen a rapid development in recent years. Compared to the 3% power conversion efficiency ($\eta$) value reported in 2007[26], to date the best OPV solar cell that employs a non-tandem bulk heterojunction (BHJ) structure has shown $\eta \sim 8.3\%$[27].

In a typical BHJ architecture (see Fig. 1a for device) a solvent-cast layer of π-conjugated polymer and fullerene-derivative blend is sandwiched between cathode and anode. The most studied polymer/fullerene blend with high $\eta$-value comprises of regio-regular poly(3-hexylthiophene) (P3HT) [donor-D] and 1-[3-(methoxycarbonyl)propyl]-1-1-phenyl)[6,6]$C_{61}$ (PCBM) [acceptor-A], of which chemical structures are shown in Fig. 1a. This blend is known to contain separate D and A nano-sized domains that facilitate both charge photogeneration and charge transport and collection in the blend[3-4]. The P3HT donor polymer absorbs in the UV-visible part of the solar spectrum which compensates the optical transparency of the fullerene molecules in the same energy range. Upon light absorption in the P3HT polymer chains, excitons are initially photogenerated in the donor nano-domains[28]; their subsequent dissociation at the D-A interfaces is facilitated by the energy level difference between the Lowest Unoccupied Molecular Orbital (LUMO) of the donor and acceptor, as well as between their Highest Occupied Molecular Orbital (HOMO) levels. To reach the D-A interfaces the excitons first diffuse towards the polymer domain boundary within ~10 ps[28-29] where upon arrival they form charge transfer excitons[30-31]. Subsequently the charge-transfer excitons separate into Coulombically-bound polaron pairs (PP), which are the intermediate species at the D-A domain interfaces having relatively long lifetime (namely few microseconds)[32]. At a later time the PP may separate into 'free' electron and hole polarons available for charge transport via the A and D domains, and can be readily collected at the anode and cathode, respectively.

One of the main challenges that OPV faces at the present time is its low $\eta$-value compared with other photovoltaic devices[1], with the recombination of PP at D-A interfaces being a major limiting factor[29,33-35]. In addition to optimizing parameters such as material mass ratio, active layer thickness, and annealing temperature; numerous other approaches have been tried to enhance the efficiency by improving the device morphology[2-5], engineering new polymer/fullerene materials[6-11], manipulating electrode property[12-15], employing tandem cell architecture[16-18], and enhancing optical absorption[19].

---

* To whom correspondence should be addressed; e-mail: val@physics.utah.edu

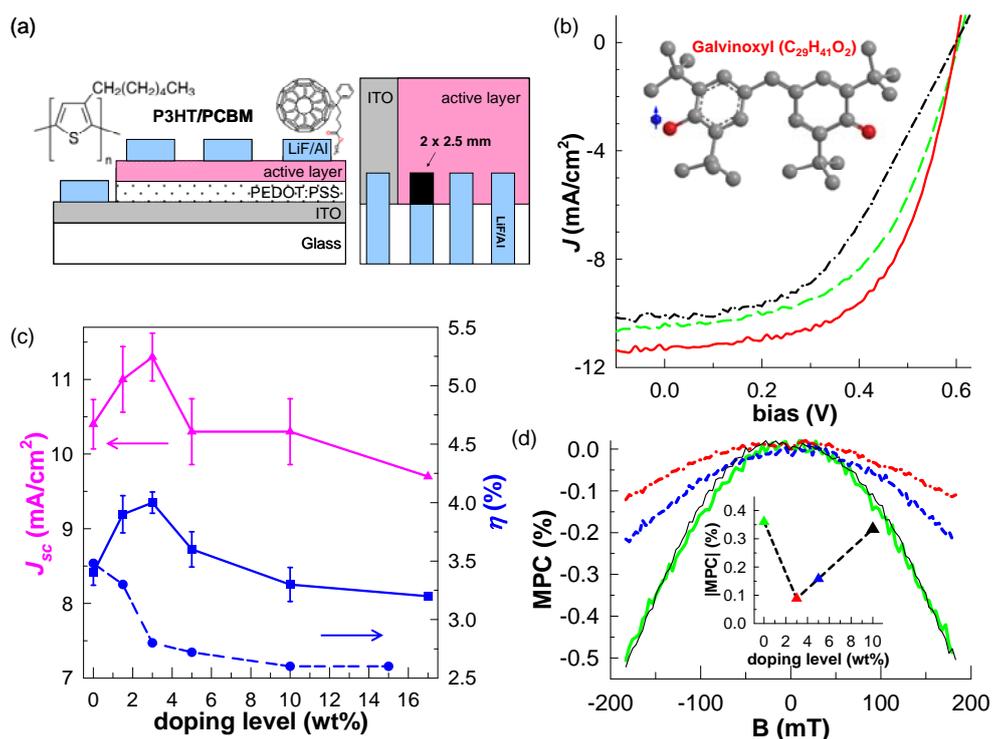

**Figure 1 Spin enhancement of OPV solar cell based on P3HT/PCBM blend doped with spin ½ Galvinoxyl radicals.** (**a**) The OPV cell structure (for fabrication see Methods). The P3HT repeat unit and PCBM molecular structure are also shown in the insets. (**b**) The I-V characteristics of OPV solar cells of pristine P3HT/PCBM blend ($\eta$=3.4%, dash), the blend doped with 3 wt% Galvinoxyl radicals ($\eta$=4.0%, solid) and the blend doped with 3 wt% precursor ($\eta$=2.8%, dash-dot) under AM1.5 'sun illumination' condition. The inset shows the Galvinoxyl molecular structure. Small sphere bisected by arrow denotes unpaired electron. (**c**) The change in OPV device properties with Galvinoxyl additive concentration; the short circuit current density, $J_{sc}$ (triangles) and power conversion efficiency, $\eta$ (squares) are shown vs. Galvinoxyl wt% in the P3HT/PCBM blend. $\eta$ of OPV devices doped with Galvinoxyl *precursor* that does not possess spin ½ radical is also shown for comparison (circles). (**d**) Magneto-photocurrent (MPC) response of OPV devices doped with Galvinoxyl (pristine: thick-solid; 3 wt%: dash-dot; 5 wt%: dash; 10 wt%: thin-solid) up to field, $B$ of 190 mT. The inset summarizes the MPC value at 150 mT vs. Galvinoxyl wt%.

A number of these approaches involve introducing nano-particles dopants (or additives) into the active layer and/or fabrication process[2-3,12-13,22-24].

In the present work we demonstrate a new method to improve OPV efficiency by doping the device active layer with spin ½ radicals to reduce PP recombination at the polymer/fullerene interfaces. We achieved a significant increase in $\eta$ by up to ~340% (in acceptor-rich devices) relative to devices based on pristine blends. We demonstrate that the spin ½ radical additives facilitate the intersystem crossing of PP from singlet to triplet spin configuration, thereby enhancing PP separation into free charges in the device; this process is unraveled via magneto-photocurrent of the doped devices. We demonstrate that the spin ½ radicals may spin flip the acceptor electron spin via an exchange mechanism that requires resonant conditions. We believe that this method may work with other D-A blends if appropriate radicals in resonance are found, in concert with other existing methods to yield even higher OPV device efficiencies.

### Results

**Galvinoxyl-doped P3HT/PCBM OPV devices.** The spin ½ radical that enhances OPV performance in this work is Galvinoxyl (2,6-di-*t*-butyl-α-(3,5-di-*t*-butyl-4-oxo-2,5-cyclohexadien-1-ylidene)-*p*-tolyloxy), a π-conjugated molecule with C$_2$ symmetry (Fig. 1b inset). The bulky *t*-butyl groups on the molecule stabilize the radical by keeping other molecules apart thus preventing further radical-radical interaction in the solid state. The unpaired electron is delocalized over the entire molecule and thus its molecular structure may be regarded as resonance hybrid of two configurations having a localized unpaired spin-polarized electron on different oxygens[36].

First we investigated the effect of Galvinoxyl doping in the active layer of a 'standard'

P3HT/PCBM device having 1.2:1 weight ratio. We note that our standard P3HT/PCBM devices were fabricated using a well-optimized recipe, and the obtained $\eta$-value is ~3.4% which is close to the published value by Plextronics[37]; it represents an upgrade from our preliminary results[38]. Fig. 1b shows that by doping 3 wt% of Galvinoxyl $\eta$ increases from 3.4% (short circuit current: $J_{sc}$=10.4 mA/cm$^2$, open circuit voltage: $V_{oc}$=0.6 V, fill factor: $FF$=0.56) to 4.0% ($J_{sc}$=11.3 mA/cm$^2$, $V_{oc}$=0.6 V, $FF$=0.62); exhibiting an 18% enhancement in the power conversion efficiency. This $\eta$ increase is significantly larger than the standard deviation in $\eta$-values of our optimal devices (±3%) (Supplementary Table S1); thus doping with Galvinoxyl unambiguously enhances the device $\eta$-value. The 8.7% increase in $J_{sc}$ that accounts for about half of the improvement in the device $\eta$, indicates that carrier photogeneration increases, carrier recombination is reduced, or both.

Fig. 1c summarizes the P3HT/PCBM (1.2:1) device properties for all investigated Galvinoxyl concentrations (1.5-17 wt%). The enhancement in $J_{sc}$ and $\eta$ induced by the Galvinoxyl radicals peaks at ~3 wt%, and gradually vanishes with further increased doping. Actually at high doping level (>10 wt%) Galvinoxyl suppresses the device performance. The optimal doping concentration (~3 wt%) at which $\eta$ maximizes divides the effect of Galvinoxyl additives into two regimes: an 'enhancement' regime, where $\eta$ increases with doping, and a 'suppression' regime, where $\eta$ decreases with doping.

We also performed magneto-photocurrent (MPC) measurements on the Galvinoxyl doped OPV devices to unravel the underlying mechanism for the increase in $J_{sc}$ with wt%. Fig. 1d shows the obtained MPC response of OPV devices having various Galvinoxyl wt%. It is clearly seen that Galvinoxyl additives reduce the MPC value without changing the field response. It has been known that MPC in OPV blends is due to magnetic field manipulation of spin triplet and singlet states within the PP species[39-41]. Therefore the reduction of MPC with wt% shows that the spin ½ radicals interfere with the intersystem crossing rates among the various spin states of the PP species, revealing the importance of the Galvinoxyl *spin* rather than its ability to serve as donor or acceptor. We therefore conclude that reduced PP geminate recombination at the D-A interfaces is responsible for the enhanced carrier photogeneration upon Galvinoxyl doping. The remaining enhancement in $\eta$ with the Galvinoxyl additives is due to an increase in $FF$ which indicates a reduced series resistance that results from improved carrier transport.

We note that the MPC reduction with Galvinoxyl wt% follows the same trend as that of the OPV enhancement with wt%. Fig. 1d inset shows that the most MPC reduction occurs at 3 wt%; the MPC response comes back to that of pristine device at 10 wt%. This further shows the existing correlation between the spin ½ properties of the Galvinoxyl additives and the OPV enhancement. In the following we describe several 'control experiments' that we performed for understanding the OPV enhancement upon Galvinoxyl doping.

In order to investigate whether the enhancement in $J_{sc}$ with Galvinoxyl doping is due to an increase in the device active layer absorption, we compared the absorption spectra of the pristine and Galvinoxyl optimally doped P3HT/PCBM films (Fig. 2a). The two spectra are very similar to each other; and, in particular the Galvinoxyl absorption peak at 430 nm (Fig. 2a), which originates from HOMO-LUMO transition, is not discerned in the doped sample absorption. We therefore conclude that the enhancement in $J_{sc}$ is not caused by a change in absorption with Galvinoxyl doping. In Fig. 2b we compare the External Quantum Efficiency (EQE) of the pristine and optimally doped devices, which is actually a better comparison than the absorption spectrum discussed above. The enhancement in EQE of the Galvinoxyl doped device does not initiate at 430 nm where the Galvinoxyl absorption is the strongest. Instead, EQE increases across the entire spectrum. We conjecture that Galvinoxyl does not act as a donor molecule in this blend system.

Morphology change and its impact on exciton diffusion towards the D-A interfaces and collection efficiency also plays an important role in determining $J_{sc}$. In order to investigate whether the film morphology changes due to the addition of Galvinoxyl molecules, we compared x-ray diffraction (XRD) patterns of the pristine and Galvinoxyl doped P3HT/PCBM films (Fig. 2c). The XRD (100) P3HT band of both films is the same. Using the XRD peak position, its full width at half maximum, and the Scherrer's relation we estimate that the P3HT domain size in both films are ~19 nm. This domain size is ideally suited to the commonly accepted 10 nm exciton diffusion length in the P3HT domains[32].

The relative intensity of 0-0 and 0-1 bands in the photoluminescence (PL) spectrum provides another way to determine the crystallization degree of the P3HT domains[42]. The normalized PL spectra of pristine and doped (3 wt%) P3HT/PCBM films are shown in Fig. 2d. The identical PL spectra indicate that the 'packing order' of polymer chains in the P3HT domains is not affected by the addition of Galvinoxyl, and thus the exciton lifetime in the P3HT domains is unchanged. Similar to the XRD and PL results, the TEM images (Supplementary Figure S1) show no observable morphology change caused by the Galvinoxyl doping. We therefore conclude that no change in film morphology is responsible for the increase in $J_{sc}$.

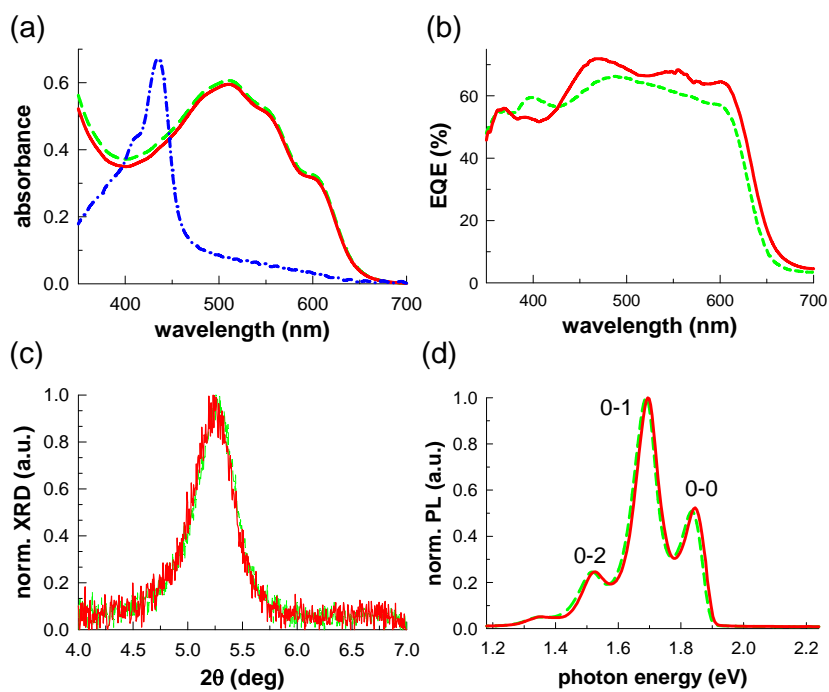

**Figure 2 Optical and structure properties of pristine and Galvinoxyl doped (3 wt%) P3HT/PCBM blend film and OPV device.** (**a**) The UV/Vis absorption spectrum of pure Galvinoxyl (dash-dot) and pristine (dash) and doped (solid) P3HT/PCBM blend. (**b**) The external quantum efficiency (EQE) spectrum of OPV solar cells based on pristine (dash) and Galvinoxyl doped (solid) P3HT/BCBM blend. (**c**) The X-Ray Diffraction (XRD) pattern of pristine (green dash) and doped (red solid) P3HT/PCBM films. (**d**) The photoluminescence (PL) spectrum of pristine (dash) and doped (solid) P3HT/PCBM. The phonon replicas are assigned.

In order to further check the importance of the Galvinoxyl spin ½ properties rather than its doping ability, we measured OPV device performance with the addition of 'Galvinoxyl precursor' molecule that has one extra hydrogen atom, and thus does not possess a spin ½ radical (its molecular structure is shown in Supplementary Figure S2) We note that the frontal orbitals are similar in the two molecules, and thus the HOMO-LUMO levels do not change much in going from Galvinoxyl to its precursor, as deduced from their similar absorption spectra (Supplementary Figure S2). In contrast to Galvinoxyl doping, we found that doping with this precursor monotonically reduces the OPV device performance (Fig. 1c). We therefore conclude that the viable mechanism for the OPV $\eta$ increase with Galvinoxyl additives is suppression of PP recombination at the D-A interfaces due to the spin ½ radicals.

**Galvinoxyl-doped P3HT-rich and PCBM-rich OPV devices.** In order to further understand the effect of Galvinoxyl doping we studied OPV cells based on P3HT-rich and PCBM-rich blends. These are: P3HT:PCBM at 2:1, 1:2 and 1:5, respectively. In P3HT-rich OPV (Fig. 3a) $\eta$ decreases monotonically with doping concentration, indicating that Galvinoxyl does not have an enhancement effect on a blend lacking sufficient PCBM acceptor molecules. In contrast, in PCBM-rich system, similar to the standard P3HT/PCBM system both enhancement and suppression regimes can be clearly identified (Fig. 3b). At high Galvinoxyl concentration (25 wt%) the doped device performance is still much better than the pristine device. In addition we notice that the enhancement effect is more significant and the optimal Galvinoxyl doping level shifts towards more heavily doping with increased PCBM content. For OPV based on P3HT:PCBM 1:2 blend, $\eta$ increases by 74% (from 1.08% to 1.88%) at optimal doping of 8 wt%; whereas for devices based on P3HT:PCBM 1:5 blend, $\eta$ increases by 336% at optimal doping of 15 wt%. In contrast, devices based on pristine PCBM alone did not show any performance enhancement upon Galvinoxyl doping. It is thus clear that Galvinoxyl enhancement of OPV devices is effective only in the presence of the P3HT/PCBM interfaces. This is consistent with our speculation that Galvinoxyl assists the PP dissociation process, which occurs exclusively at the D-A domain interfaces.

**Galvinoxyl-doped OPV devices with other donors and acceptors.** We carried out additional studies of

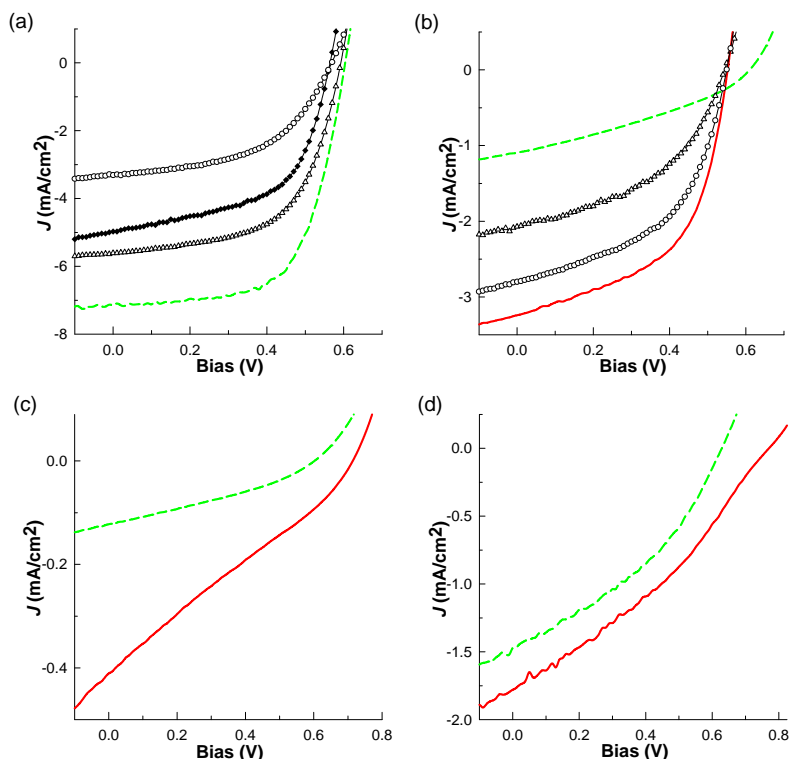

**Figure 3 I-V characterization of OPV devices based on pristine and Galvinoxyl doped donor/acceptor blends. (a)** Polymer donor rich P3HT/PCBM 2:1 blend at various Galvinoxyl doping levels: pristine ($\eta$=2.8%, dash), 1.5 wt% ($\eta$=1.9%, triangles), 3 wt% ($\eta$=1.6%, diamonds) and 10 wt% ($\eta$=1.0%, circles). **(b)** Fullerene acceptor rich P3HT/PCBM 1:5 blend at various Galvinoxyl doping levels: pristine ($\eta$=0.22%, dash), 10 wt% ($\eta$=0.52%, triangles), 15 wt% ($\eta$=0.96%, solid) and 25 wt% ($\eta$=0.77%, circles) **(c)** Regio-random P3HT/PCBM 1.2:1 blend at different Galvinoxyl doping levels: pristine ($\eta$=0.02%, dash), 3 wt% ($\eta$=0.08%, solid). **(d)** MEH-PPV/PCBM 1:1 blend at different Galvinoxyl doping levels: pristine ($\eta$=0.26%, dash), 17 wt% ($\eta$=0.34%, solid).

Galvinoxyl doping OPV devices fabricated using blends of other donor and acceptor materials, to explore whether the enhancement effect is limited to the P3HT/PCBM system. We replaced the donor polymer in our standard P3HT/PCBM blend with regio-random P3HT (RRa-P3HT), which has a similar chemical structure to regio-regular P3HT except that the hexyl side-chains are arranged randomly. In Fig. 3c we show that in Galvinoxyl doped (3 wt%) RRa-P3HT/PCBM device, $J_{sc}$ improves from 0.12 to 0.41 mA/cm$^2$. Coupled with $V_{oc}$ increase from 0.61 V to 0.72 V, a 300% increase in $\eta$ is observed. This is not surprising considering that the dominant loss mechanism in the RRa-P3HT/PCBM blend is PP recombination at the D-A interfaces[43], a process that Galvinoxyl doping is expected to directly suppress[44]. Similar result was obtained when we replaced the donor polymer with poly[2-methoxy-5-(2-ethyl-hexyloxy)-1,4-phenylene-vinylene] (MEH-PPV). In Fig. 3d we show that $J_{sc}$ of the doped (17 wt%) device increases from 1.47 to 1.78 mA/cm$^2$. With $V_{oc}$ increasing from 0.63 V to 0.78 V, a 30% increase in $\eta$ is observed in devices based on this blend.

In contrast, attempts to enhance $\eta$ were unsuccessful in OPV devices based on blends with acceptor fullerenes other than PCBM. When we replaced PCBM with bis-PCBM, a bisadduct analogue of PCBM; or Indene-C$_{60}$ bisadduct (ICBA) that is a high LUMO level fullerene[9], we found that $\eta$ decreases monotonically with Galvinoxyl doping. It is worth mentioning that when we replaced a fraction of the PCBM acceptor in the blend with bis-PCBM, the enhancement effect was still observable. We thus conclude that PCBM in the organic blend is crucial for effective Galvinoxyl doping enhancement.

Other radials that we tested but did not show notable enhancement effect in P3HT/PCBM OPV system are: (2,2,6,6-Tetramethylpiperidin-1-yl)oxyl (TEMPO), $\alpha,\gamma$-Bisdiphenylene-$\beta$-phenylallyl (BDPA) and bis(1-hydroxy-2,2,4,6,6-pentamethyl-4-piperidinyl) oxalate (Supplementary Figure S3, Table S2 and Supplementary Methods).

## Discussion

The experimental evidence indicates that the cause for the enhanced η-value in the Galvinoxyl doped OPV devices is the reduced PP recombination rate at the P3HT/PCBM domain interfaces due to the spin ½ radical additives. We still need to unravel the mechanism by which Galvinoxyl reduces PP recombination. Since Galvinoxyl is a spin ½ radical, we propose a 'spin-flip' mechanism that facilitates PP separation at the P3HT/PCBM interfaces by converting photogenerated PP from spin singlet to triplet (Fig. 4a) via spin exchange interaction between the PP and Galvinoxyl. Since triplet PP has longer lifetime than singlet PP, the enhanced intersystem crossing results in a longer-lived species having a better chance to dissociate.

Consider that a photogenerated exciton in the P3HT domain has spin-up electron in the LUMO level and spin down hole in the HOMO level. Upon arrival at the D-A interface the electron transfers to the PCBM LUMO level, forming a singlet PP (Fig. 4a upper left) with the spin-down hole in the P3HT HOMO level. The singlet PP species can either dissociate into free carriers (polarons) in the P3HT and PCBM separate domains, or geminately recombine. When a spin ½ radical such as Galvinoxyl, which has spin-polarized singly occupied molecular orbital (SOMO) and LUMO levels with designated spin orientations is introduced next to the singlet PP, then it may form a complex with PCBM providing a spin-down polarized empty LUMO level *in resonance* with the spin-up filled PCBM LUMO level next to the charged PCBM molecule (Fig. 4a right). This mediates an exchange interaction between the up-spin negative polaron and the 'virtual' down-spin of Galvinoxyl LUMO that flips the polaron up-spin to down-spin, thereby forming a lower-energy triplet PP (Fig. 4a bottom left). The PP triplet species has a longer lifetime because it is 'spin-forbidden', having a reduced geminate recombination rate. This may facilitate its dissociation into free polarons. The same mechanism can be equally applied for a photogenerated PP with spin-down electron in the P3HT LUMO level, via exchange interaction with a spin-up defined LUMO level of Galvinoxyl radical.

We conclude that Galvinoxyl additives activate a spin-flip process that converts PP species at the D-A interfaces from spin singlet to triplet, and this reduces the overall PP recombination rate. Since our experimental evidence indicates that this process is effective only with the presence of PCBM, and is more effective with increasing PCBM concentration we therefore believe that Galvinoxyl forms a complex with the PCBM at the P3HT/PCBM interfaces. Also the OPV enhancement is maximum at certain Galvinoxyl optimal concentration; this can be understood if nearest neighbor Galvinoxyl molecules at high wt% are spin-paired to form spin singlet. Therefore overdose of Galvinoxyl molecules may reduce their ability to provide the spin-flip mechanism necessary for reducing the PP recombination rate.

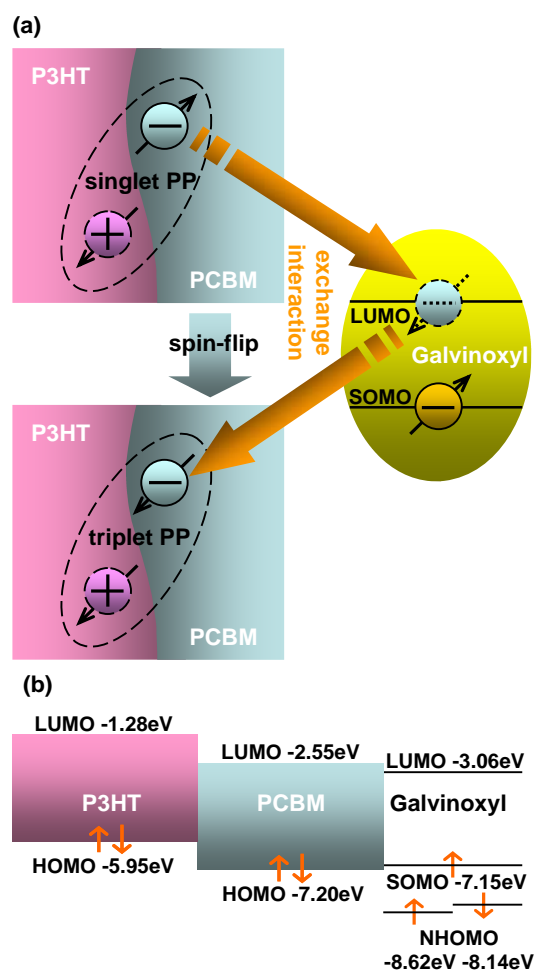

**Figure 4 Spin-exchange mechanism and DFT calculation.** (**a**) The spin exchange mechanism, where the photogenerated polaron-pair (PP) at the D-A domain interface changes its spin configuration from singlet to triplet augmented by the Galvinoxyl spin ½ radical. (**b**) The calculated HOMO, LUMO and SOMO levels of P3HT, PCBM and Galvinoxyl that show a clear resonance between the radical and acceptor LUMO levels.

For an effective spin exchange interaction, energy resonance among the charge PCBM and Galvinoxyl molecular levels is crucial. To investigate this constraint we performed DFT calculations of the ground-state frontier molecular orbital levels of P3HT, PCBM and Galvinoxyl using the CAM-B3YP functional. The level alignments among P3HT,

Galvinoxyl and PCBM are shown in Fig. 4b. P3HT LUMO level (-1.28 eV) is substantially higher than PCBM LUMO level (-2.55 eV), and this facilitates the initial charge transfer that results in PP formation at the D-A interfaces. Importantly we found that PCBM LUMO (-2.55 eV) is within resonance of the Galvinoxyl LUMO (-3.06 eV), and this helps mediating the exchange interaction of the spin flip process. In addition the formation of Galvinoxyl/PCBM complex may also be important for the spin flip process. We note that the calculated energy levels are for isolated individual molecules, which may change in forming the complex due to inter-molecular interaction. Nevertheless our calculations show that the energy resonance between Galvinoxyl and PCBM is noticeably better than those between TEMPO and PCBM, or between Galvinoxyl and other acceptors such as bis-PCBM and ICBA (Supplementary Figure S4). This is consistent with our experimental observations that these alternative D-A blends show no significant OPV $\eta$ enhancement.

Finally we note that spin ½ additives may also enhance the electroluminescence (EL) efficiency in organic light emitting diodes (OLED). The ratio of singlet/triplet PP in OLED is usually constrained to be 1:3 due to spin statistics dictated by Quantum Mechanics. Thus the number of singlet PPs that are precursors to singlet excitons that ultimately emit EL is severely limited. This is especially an acute problem for OLEDs with blue emission that relies on singlet excitons to yield EL; in particular supplying the missing blue color for white OLEDs. By adding spin ½ radicals to the active layer we may change the singlet/triplet statistical limit, since the radical may provide another intersystem crossing channel that may increase the singlet PP yield on the expense of triplet PP. This may be a viable route of future investigations, and our results may show the way to advance this research direction because we emphasize the need of energy resonance in the spin exchange mechanism between the spin ½ additives and the polaron levels in the active layer.

## Methods

**OPV device fabrication.** The bulk heterojunction OPV devices investigated in this study were composed of a transparent indium tin oxide (ITO) anode; a spin-cast polyethylenedioxythiophene/polystyrene sulphonate (PEDOT/PSS) hole transport layer; an active material layer spin-cast from a blend of polymer donor, fullerene acceptor and spin ½ radical Galvinoxyl (when applicable); and capped with LiF/Al cathode, as shown in Fig. 1a. The ITO-coated glass substrates (Delta Technology, CB-50IN) were cleaned by ultrasonic treatment (in acetone, detergent, deionized water and methanol sequentially) and oxygen plasma treatment. The PEDOT/PSS (Clevios, P VP AI 4083) layer was spin-cast at 5000 RPM for 20 sec at ambient conditions, and transferred to a nitrogen-filled glovebox ($O_2$ <1 ppm) for annealing at 120 ºC for 30 min. The organic blend that yielded the best device performance ($\eta$~4%) comprised of P3HT (Plextronics, Plexcore OS 2100), PCBM (purity >99.9%) and Galvinoxyl (Aldrich, G307). It was prepared in the following way: P3HT (16 mg/ml) and PCBM were dissolved at 1.2:1 weight ratio in 1,2-dichlorobenzene (ODCB). The blend was heated at 50 ºC for 30 min and stirred over night before mixing with Galvinoxyl (3 wt%, defined as percentage of total P3HT/PCBM weight) and stirring for 1 additional hr. Other P3HT/PCBM blends were prepared by changing the D-A weight ratio while keeping the total P3HT/PCBM mass unchanged. The active layer was spin-cast from the blend at 400 RPM for 6 min and annealed at 150 ºC for 30 min. The device fabrication was completed by thermally evaporating a 1 nm thick film of LiF followed by a 100 nm thick film of Al. Each device was 2 mm×2.5 mm in dimension, as shown in Fig. 1a. Finally the completed device was encapsulated under a cover glass using UV-curable optical adhesive (Norland, NOA 61). Devices using other donor/acceptor materials were fabricated using the same protocol. The PCBM-only device film was spin-cast from a 20 mg/ml chlorobenzene solution at 400 RPM and baked at 50 ºC for 2 hr. The MEH-PPV/PCBM (1:1 weight ratio) film was spin-cast from a 3 mg/ml toluene solution at 2000 RPM and annealed at 120 ºC for 30 min.

**OPV device characterization.** Device I-V characteristics were measured using a Keithley 236 Source-Measure unit. The light intensity of the solar simulator, composed of a xenon lamp and an AM1.5G filter, was calibrated at 100 mW/cm$^2$ using a pre-calibrated silicon PV cell. Monochromatic lights used to measure device EQE characteristics were modulated at 36 Hz and filtered by a KG5 filter before sending into the photodetector or OPV device. The photodetector (OPV) device output current was measured using phase sensitive lock-in technique.

**Magneto-photocurrent measurements.** When measuring magneto-photocurrent (MPC), the OPV devices were transferred to a cryostat that was placed in between the two poles of an electromagnet producing magnetic field, $B$ up to ~200 mT. The devices were illuminated with a tungsten lamp and measured at zero bias using a Keithley 236 apparatus, while sweeping the external magnetic field. The MPC is defined as MPC($B$)=[PC($B$)/PC(0) -1].

**Other characterization methods**. The absorption spectrum was measured using a UV/Vis spectrophotometer (PerkinElmer Lambda 750) at ambient. X-ray diffraction (XRD) was obtained using the Cu K$_\beta$ line and collected by a diffractometer (Philips X'Pert PRO) in powder diffraction mode. Photoluminescence emission was excited by 100 mW continuous wave laser at 488 nm. The emitted light was collected using a spherical mirror and dispersed by an Acton 300 spectrometer before sending to a photodetector. Transmission Electron Microscopy (TEM) images were taken under defocused (-25 μm) condition, which is reported to provide enhanced contrast between the P3HT and PCBM regions. Films used for absorbance, XRD and TEM measurements were prepared in the same way as the active layer of the actual OPV device. Films used for PL measurements were drop-cast from 16mg/ml P3HT solution and annealed at 150 ºC for 30 min.

**Density functional theory (DFT) calculation.** The ground state molecular orbital levels were calculated by the Gaussian 09 program[45] with density functional theory (DFT, CAM-B3LYP/6-311g**). The molecular geometry was fully optimized at CAM-B3LYP/6-31g* level[46-47]. The DFT calculation for P3HT were performed with periodic boundary conditions[48] (see also Supplementary Methods).

## Acknowledgements
This work was supported in part by the DOE Grant No. DE-FG02-04ER46109 (TEMPO radical additive and the initial Galvinoxyl studies), NSF-MRSEC program Grant No. DMR-1121252 (model calculations and MPC studies), and NSF Grant No. DMR-1104495 (final Galvinoxyl studies).

## Author contributions
Y.Z., T.P.B. and D.J.M. were responsible for fabricating and characterizing the OPV devices; B.R.G. was responsible for the MPC measurements; X.Y. and F.L. were responsible for the DFT calculation; and Z. V. V. was responsible for the project planning, group managing, and writing the manuscript final version.

## Additional information
Supplementary Information accompanies this paper on www.nature.com. Correspondence and requests for materials should be addressed to Z.V.V.